\documentclass[12pt,epsf]{iopart}
\usepackage [dvips]{graphicx}
\usepackage{iopams}
\usepackage{epsfig}
\usepackage{verbatim}

\addtocounter{secnumdepth}{1}
\newcommand{\tfrac}[2]{\mbox{$\frac{#1}{#2}$}}
\font\capfont=cmbx12 at 50 pt 
\newbox\capbox \newcount\capl \def\a{A}
\def\docappar{\medbreak\noindent\setbox\capbox\hbox{%
\capfont\a\hskip0.15em}\hangindent=\wd\capbox%
\capl=\ht\capbox\divide\capl by\baselineskip\advance\capl by1%
\hangafter=-\capl%
\hbox{\vbox to8pt{\hbox to0pt{\hss\box\capbox}\vss}}}
\def\cappar{\afterassignment\docappar\noexpand\let\a }

\begin{document}

\newcommand{\ee}{{\rm e}}
\newcommand{\dd}{{\rm d}}
\newcommand{\p}{\partial}
\newcommand{\calT}{T^{\rm in}}
\newcommand{\bJ}{\bar{J}}
\newcommand{\bj}{\bar{j}}
\newcommand{\Jfree}{J^{\rm F}}
\newcommand{\Jjam}{J^{\rm J}}

\newcommand{\Jin}{J^{\rm F}}
\newcommand{\Jout}{J^{\rm J}}

\newcommand{\Jinc}  {J^{\rm{F}}_{\rm c}}
\newcommand{\Joutc}{J^{\rm{J}}_{\rm c}}

\newcommand{\rhoin}{\rho^{\rm{F}}}
\newcommand{\rhoinc}{\rho^{\rm{J}}_{\rm c}}
\newcommand{\rhoout}{\rho^{\rm{J}}}
\newcommand{\rhooutc}{\rho^{\rm{F}}_{\rm c}}

\newcommand{\ac}{a_{\rm c}}
\newcommand{\alphac}{\alpha_{\rm c}}
\newcommand{\hP}{\hat{P}}
\newcommand{\hT}{\hat{T}}
\newcommand{\htau}{\hat{\tau}}

\newcommand{\la}{\langle}
\newcommand{\ra}{\rangle}
\newcommand{\beq}{\begin{equation}}
\newcommand{\eeq}{\end{equation}}
\newcommand{\bea}{\begin{eqnarray}}
\newcommand{\eea}{\end{eqnarray}}
\def\lsim{\:\raisebox{-0.5ex}{$\stackrel{\textstyle<}{\sim}$}\:}
\def\gsim{\:\raisebox{-0.5ex}{$\stackrel{\textstyle>}{\sim}$}\:}


\thispagestyle{empty}
\title[Frozen shuffle update for open TASEP]{
Frozen shuffle update\\
for an asymmetric exclusion process
\\
with open boundary conditions
}

\author{{C. Appert-Rolland, J. Cividini, and H.J.~Hilhorst}}

\address{1 - University Paris-Sud; Laboratory of Theoretical Physics\\
B\^atiment 210, F-91405 ORSAY Cedex, France.}

\address{2 - CNRS; UMR 8627; LPT\\
B\^atiment 210, F-91405 ORSAY Cedex, France.}

\ead{\mailto{Cecile.Appert-Rolland@th.u-psud.fr}, 
\mailto{Julien.Cividini@th.u-psud.fr},
\mailto{Henk.Hilhorst@th.u-psud.fr}}

\begin{abstract}
\noindent 
We introduce a new update algorithm for exclusion
processes, more suitable
for the modeling of pedestrian traffic.
Pedestrians are modeled as hard-core particles
hopping on a discrete lattice, and are updated in a fixed
order, determined
by a {\em phase} attached to each pedestrian.
While the case of periodic boundary conditions
was studied in a companion paper, we consider here
the case of open boundary conditions.
The full phase diagram is predicted analytically
and exhibits a transition between a free flow phase 
and a jammed phase.
The density profile is predicted in the frame
of a domain wall theory, and compared to Monte Carlo
simulations, in particular in the vicinity of
the transition.\\

\noindent
{{\bf Keywords:} pedestrian traffic, exclusion process, shuffle
  update, boundary driven phase transition}
\end{abstract}
\vspace{12mm}

\noindent LPT Orsay 11/54

\maketitle
\newpage


\section{Introduction} 
\label{sect_introduction}

\cappar
As part of the effort to describe pedestrian motion,
the past years have seen the development of
cellular automata based models, among which
the so-called `floor field' model \cite{burstedde01b,schadschneider02}.
These models represent pedestrians as particles
that jump from site to site on a discrete lattice,
with an exclusion principle that forbids two pedestrians
to simultaneously occupy the same site.
In general,
the pedestrians' positions are subject to a parallel update procedure,
{\it i.e.,} the particles attempt to jump at the same instants of time.
Although this type of update ensures a quite regular motion 
under free flow conditions, it creates conflicts
when two or more pedestrians try to move at the same
time to the same target site.

Introducing an arbitrary numerical decision procedure to resolve these
conflicts can be considered as a drawback and
alternative update procedures have been looked for.
In reference \cite{klupfel07}, for example,
large pedestrian simulations are performed
by means of a random shuffle update. In this sequential 
update scheme an updating order 
for the pedestrians is drawn at random at the beginning of each time step. 

The random shuffle update was first proposed and studied
in \cite{wolki_s_s06,wolki_s_s07,smith_w07a}. It was applied there
to the Totally Asymmetric Simple Exclusion Process (TASEP),
which can be seen as a basic model for pedestrian traffic.
More generally, exclusion processes are archetypes
of out-of-equilibrium systems and have been used
to model various transport phenomena, including
road and intracellular traffic, or ant motion
\cite{chowdhury_s_n05}.

Here and in a companion paper \cite{appert-rolland_c_h11a}
we introduce a new variant of the shuffle update
called the {\em frozen shuffle update}.
Under this updating procedure
each particle $i$ is assigned, either at the beginning of the
simulation or when it enters the system,
a random {\em phase} $\tau_i\in[0,1)$.
In each time step particles are updated in
the order of increasing phases.
The phases may be thought of as the phases in the
walking cycle, where one pedestrian may be 
are slightly in advance with respect to another one.
This advance then
determines who passes first in case of conflict.

We have recently studied
the frozen shuffle update for the TASEP on a ring,
i.e. on a closed system \cite{appert-rolland_c_h11a}.
This system is appropriately described by its current $J$ as
a function of the particle density $\rho$.
We found \cite{appert-rolland_c_h11a}
that the TASEP on a ring undergoes a phase transition at
a critical density $\rho=\tfrac{2}{3}$, where
$J(\rho)$ attains its maximum through a linear cusp.
The critical point separates a `free flow' phase from a `jammed' phase.

In the present paper we 
study an open-ended TASEP with frozen shuffle update.
Whenever a new particle, say $i$, enters through the open boundary, 
its place in the updating order is specified by a newly drawn $\tau_i$.
It will then keep this same $\tau_i$ until it
leaves the system. In section \ref{sect_modeldef}
we define the frozen shuffle update. 
We will consider only its deterministic version in which allowed jumps
are carried out with probability one.
In addition to the bulk rules of motion,
which are the same as on the ring \cite{appert-rolland_c_h11a},
we specify in section \ref{sect_rules} the rules by which particles 
enter and leave the system.
This procedure will depend on two additional parameters,
an entrance probability $\alpha$ and an exit probability $\beta$.
In section \ref{sect_phase} the phase diagram of the system
in the $\alpha\beta$ plane will be determined analytically. 
We find a transition line $\alpha=\beta$ between, again, a free flow 
and a jammed phase. The bulk particle density
$\rho(\alpha,\beta)$ and the current $J(\alpha,\beta)$ 
are determined analytically and appear
in excellent agreement with simulations.
In section \ref{sect_dw}
the finite size effects, in particular near the transition line, are
described by a domain wall approach.


\section{Frozen shuffle update for the 1D TASEP}
\label{sect_modeldef}


\subsection{Frozen shuffle update}
\label{sect_definition}

We consider a system of hard-core particles
({\it i.e.} at most one per lattice site)
on a one-dimensional lattice of $L$ sites
numbered from left to right by $k=1,2,\ldots,L$. 
The particles enter the system at $k=1$,
make hops of a single lattice distance to the right,
and leave the system at $k=L$.
The hops are executed according to the 
following {\em frozen shuffle\,} update scheme.
A particle $i$, when it enters the system, is assigned
a random phase $\tau_i\in[0,1)$, which it keeps until 
it leaves the system.
The phases are drawn from a predefined distribution.
In each time step all particles are updated ({\it i.e.}
attempt to hop one step to the right) in the order
of increasing phases.
An attempted hop is successful if and only if the target 
site is empty.

Throughout, the variable $t$ will stand for continuous time; 
time steps will be indicated by integer values $t=0,1,\ldots,s,\ldots$.
The particle configuration at time $t=s$ is considered to be the result of the
$s$th time step. 


\subsection{Two stable bulk states}
\label{sect_stablestates}

In recent work \cite{appert-rolland_c_h11a} 
we have identified
two distinct stable bulk states sustained
by this update scheme.
In an infinite homogeneous system
these may be discussed without reference to boundary conditions.
They are the {\em free flow\,} state and the {\it jammed\,} state. 

{\it Free flow state.\,\,}
In the free flow state all particles are able to
move forward at each time step. 
Hence they all have a velocity $v=1$, that is, one lattice distance 
per time step.
The current $\Jfree=\rho v$ in the free flow state 
is therefore in these units given by
\begin{equation}
\Jfree = \rho\,.
\label{dJfree}
\end{equation}
The technical definition is that {\it the system is in a free flow state if
whenever two adjacent lattice sites are occupied by particles, the
particle to the right has the lower phase}.

{\it Jammed state.\,\,}
The second stable bulk state is the {\em jammed} state.
This is a highly compacted state of a special kind.
Technically, {\it the system is in the jammed state
if, whenever a particle has a smaller (larger) phase
than its follower, then it is located
on the site immediately to the right of its follower
(is separated from its follower by at most a single hole)}.
As a consequence, in this state,
a fraction of the particles remains
blocked at each time step; 
the average particle velocity therefore is less than unity.

\begin{figure}
\begin{center}
\scalebox{.55}
{\includegraphics{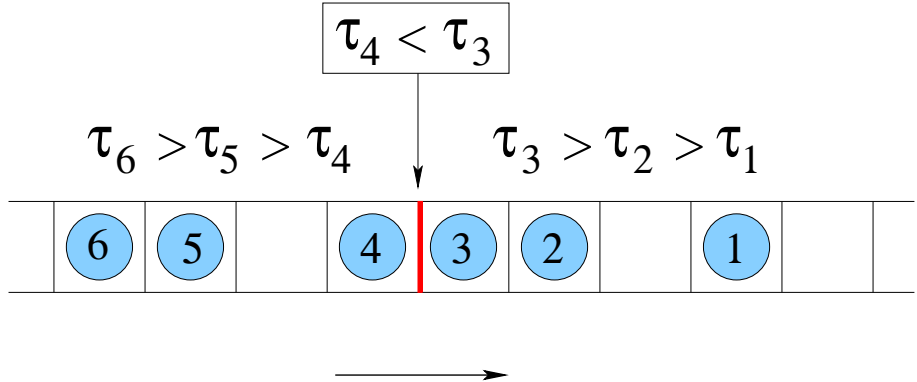}}
\end{center}
\caption{\small Six hard-core particles on a one-dimensional lattice
whose sites are represented by square boxes.
The arrow indicates the hopping direction.
Our convention will be to place a heavy (red) vertical bar
to the left of each particle that has a larger phase than
its follower (in the present example, particle 3).
The configuration shown is not a free flow state, since the
particle pair $(3,4)$ violates the free flow condition; particle 4 is
unable to move during the upcoming time step.
The configuration is not a jammed state either, since the holes between
the particle pairs $(1,2)$ or $(4,5)$ violate the jammed state condition.
}
\label{fig_flox}
\end{figure}

{\it Stability.\,\,}
The  conditions for the free flow state and the jammed state
are illustrated by figure \ref{fig_flox}.
It is tacitly understood that these conditions are to be evaluated 
after completion of a full time step.
There exist many particle configurations that are neither free flow
nor jammed states%
\footnote{There also exists a small class of configurations that are,
technically, both `free flow' and `jammed', namely the configurations
in which {\em each} platoon is preceded
by a hole; this fact is of little consequence for the discussion.}. 
For a system on a ring lattice
the latter two were shown \cite{appert-rolland_c_h11a}
to be attractors in phase space
toward which arbitrary initial states evolve.
They are stable in the sense that once the system has
entered one of them, it will stay in it. 

{\it Platoons.\,\,} 
We now look for the counterpart in the jammed phase
of the current-density relation (\ref{dJfree}).
The flow in the jammed state is most easily discussed in terms of platoons.
A {\it platoon\,} is a succession of particles occupying 
consecutive sites and having increasing phases;
the first particle of a platoon having a lower phase than the particle
that precedes it; and
the last particle of a platoon having a higher phase than the
particle that follows it.
It was shown \cite{appert-rolland_c_h11a}
that in the jammed phase
each particle belongs to a platoon.

Frozen shuffle update leads to simple platoon dynamics.
A platoon moves as a whole, {\it i.e.,} if its first particle can move,
then in the same time step
the whole platoon will move one lattice distance to the right.
Equivalently, one may say that the hole initially in front of the platoon
has hopped across it to the left over a distance equal
to the platoon length (as will be further illustrated in
section \ref{sect_exitingcurrent}).
This allows us to make the following statement 
\cite{appert-rolland_c_h11a}. 
Let $\nu$ be the average platoon length in some suitable spatially
homogeneous statistical ensemble.
Since the hole density is $1-\rho$,
and since only platoons that have a hole in front of them
can move, the current $\Jjam$ in a jammed phase is given by 
\begin{equation}
\Jjam = \nu (1-\rho)\,,
\label{dJjam}
\end{equation}
which is the relation that we looked for. Henceforth the superscripts
${\rm J}$ and ${\rm F}$
will indicate quantities referring to the jammed and the free flow
state, respectively.

As the final comment of this section we remark that
expressions (\ref{dJfree}) and (\ref{dJjam}) refer to
spatially homogeneous states 
of, in principle, infinite extent.
Our interest below will be
in a finite system with boundary conditions specified
by two parameters $\alpha$ and $\beta$.
In the next section we define these boundary conditions
and express the stationary state current $J(\alpha,\beta)$ 
of this finite system in terms of the
currents $\Jfree$ and $\Jjam$ discussed above.


\section{Rules for entering and exiting}
\label{sect_rules}


\subsection{Entering rule}
\label{sect_enteringrule}

\begin{figure}
\begin{center}
\scalebox{.55}
{\includegraphics{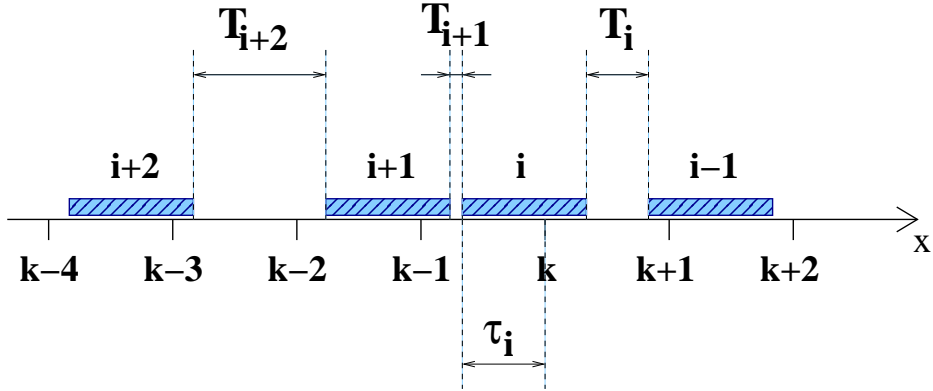}}
\end{center}
\caption{\small 
Rods move at constant speed $v=1$
from left to right along the $x$ axis.
The time interval between the tail of the $i$th and the head of the 
$(i+1)$th rod is called $T_i$\,. The $T_i$ are drawn
randomly from the exponential distribution (\ref{defPT}).
We have also represented the underlying discrete lattice,
that allows to define the mapping between the continuous
and the discrete model.
If the snapshot is taken at an integer time $s$, then
the phase $\tau_i$ of particle $i$ in the discrete model
corresponds in the continuous model to the distance
between the tail of rod $i$ and the integer position $k$
that rod $i$ overlaps
(since $v=1$, time and space may be identified).
}
\label{fig_rods}
\end{figure}

The rule governing the entering of particles
into the system is motivated by
an equivalent hard rod description \cite{appert-rolland_c_h11a}
of the system in the free flow state%
\footnote{This equivalence breaks down when there is jamming,
as it would lead to overlapping rods.}:
Let nonoverlapping rods of unit length, as shown in figure
\ref{fig_rods}, move in continuous time at constant speed $v=1$
along the $x$ axis.
At any integer instant of time $s$,
each rod $i$ will cover exactly one integer spatial coordinate $x=k$.
We will say that at time step $s$
the particle $i$ associated with the rod $i$
occupies lattice site $k$ 
in the corresponding discrete model.
This particle $i$ will then be
on lattice site $k+1$ at time $s+1$, and so on.
The (noninteger) time $s+\tau_i$ at which rod $i$
uncovers position $k$ and starts to cover position $k+1$
defines the phase $\tau_i$ of particle $i$ in the discrete model.

If the rods are (at any instant of time) uniformly distributed
in space, subject only to the no-overlap constraint,
then the intervals $T_i$ between them
(see figure \ref{fig_rods}) are independent variables 
drawn from an exponential distribution
\begin{equation}
P(T) = a {\rm e}^{-aT}, \qquad T\geq 0.
\label{defPT}
\end{equation}
where $a$ is a positive parameter.

The equivalence between the rod and the particle motion
lead us to adopt the following rule for particles 
entering the lattice, in order to have the same
time interval distribution between particles
in the discrete system.
A particle can be injected from the exterior onto
the first lattice site (labeled 1) only if that site is empty.
If site 1 gets emptied at some instant of time $s + \tau_{i-1}$
(due to particle $i-1$ moving forward to site 2),
then it will be reoccupied by a new particle $i$
at a random time $t_{i}=s + \tau_{i-1} + T_i$ with $T_i$ 
distributed according to (\ref{defPT}).
This new particle will therefore have a phase $\tau_i$ given by the
recurrence relation
\bea
\tau_i &=& t_i\mbox{ mod }1 \nonumber\\[2mm] 
       &=& (\tau_{i-1}+T_i)\mbox{ mod }1,
\label{reltaui}
\eea
to which we will return in section \ref{sect_rules}.

Below it will sometimes be convenient
to use the probability,
to be called $\alpha$, 
that the initial site, when empty at some arbitrary (real) instant of
time $t$, is occupied at time $t+1$.
This quantity is  independent of $t$ and
with (\ref{defPT}) it follows that 
\bea
\alpha &=& 
\mbox{Prob}\{T<1\} \nonumber\\[2mm]
&=& 1-{\rm e}^{-a}.
\label{defalpha}
\eea
We will refer to $a$ and $\alpha$ as the `entrance rate' and `entrance
probability', respectively.


\subsection{Entering current}
\label{sect_enteringcurrent}

Let $\rhoin$ and $\Jin$ denote the spatial density and the current,
respectively, of the incoming particles under free flow conditions. 
They then satisfy the free flow relation (\ref{dJfree}), that is,
$\Jin=\rhoin$. 
We now express these quantities in terms of the parameter $a$.
Let $\calT_N$ be the time needed for $N$ particles $i=1,2,\ldots,N$
to enter the system. Then $\calT_N$ is the sum of the time intervals
$T_i$ and, for each particle, of a unit time interval during which it
blocks the access of a new particle to site 1.
In the limit $N\to\infty$ the stochastic variable $\calT_N$ becomes sharply
peaked around its average, that is,
\beq
\lim_{N\to\infty} \frac{\calT_N}{N} =
\lim_{N\to\infty} \frac{1}{N}\sum_{i=1}^N(T_i+1) 
=\overline{T_i}+1 = \frac{1}{a}+1. 
\label{calcJ}
\eeq
Since $\Jin=\lim_{N\to\infty} N/\calT_N$\,, 
it follows that at the left hand end of the 
lattice we have, {\it under free flow entrance conditions,}
\beq
\Jin=\frac{a}{1+a}\,, \qquad \rhoin=\frac{a}{1+a}\,.
\label{xJin}
\eeq
The injection procedure seeks therefore to impose 
onto the system a free flow state
with a density and current (\ref{xJin})
characterized by the parameter $a$ (or $\alpha$).
This free flow is interrupted only when a particle
meets with an obstacle: that might be a barrier 
at the exit point or a jam caused by other particles
anywhere in the interior of the system. In the special case that
site 1 is occupied by a particle which is itself
blocked, the entering rule of section \ref{sect_enteringrule} still
applies, but in that case the free inflow stops. 
Before commenting further we will now examine what happens at the
exit point.


\subsection{Exiting rule}
\label{sect_exitingrule}

We choose to adopt the following rule for particles leaving the system.
If the last lattice site, labeled $L$, 
is occupied by a particle $i$ with phase $\tau_i$,
then, at each instant of time $s+\tau_i$,
particle $i$ leaves the system with probability $\beta$ 
and stays on site $L$ with probability $1-\beta$.
For $\beta=1$ the free flow is unimpeded at the exit. 
However, when $\beta<1$, the free flow will at times 
be randomly impeded, which may create a slowly advancing
waiting line for the particles near the exit.
When the rear end of this waiting line propagates all the way backward to the
entrance, the system is in its jammed state. 
Below we will investigate the conditions for this to happen.


\subsection{Exiting current}
\label{sect_exitingcurrent}

Let the particles near the exit point be in a jammed state
configuration and let
$\rhoout$ and $\Jout$ denote their density and their current,
respectively, under this condition.
These two quantities
then satisfy the jammed state relation (\ref{dJjam}), that is, 
$\Jout=\nu(1-\rhoout)$.
We will now show how they are determined by the 
parameters $\beta$ and $\alpha$.

\begin{figure}
\begin{center}
\scalebox{.55}
{\includegraphics{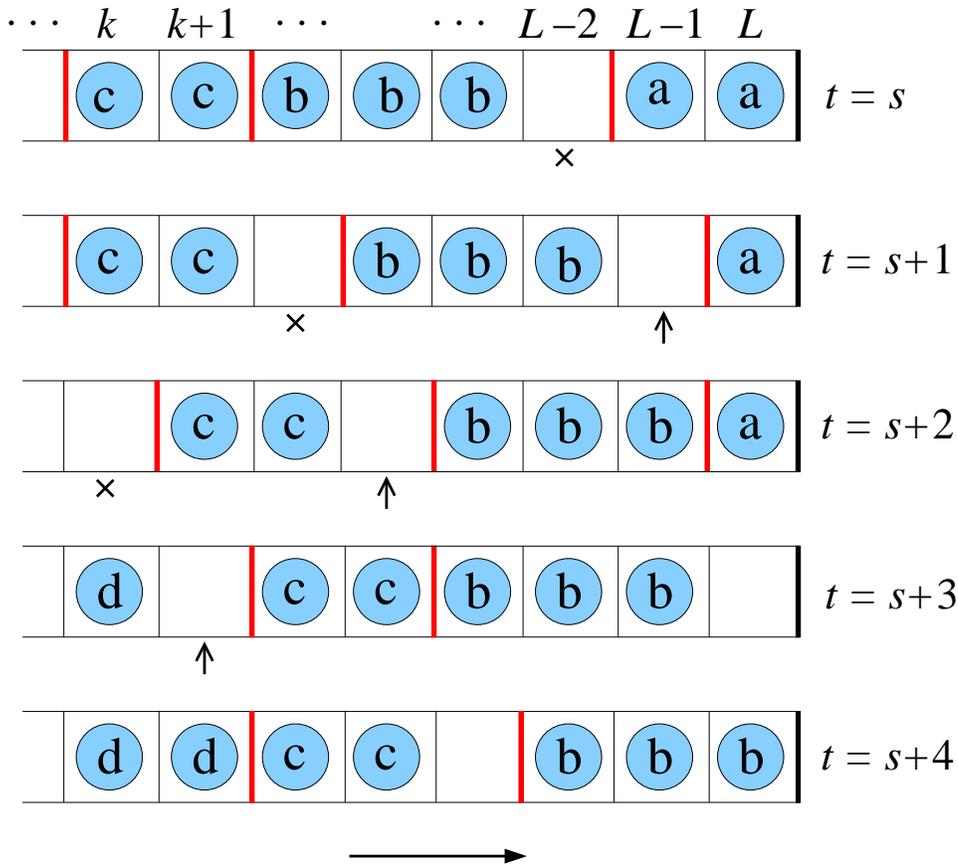}}
\end{center}
\caption{\small The lattice near the exit site $L$, 
shown at five successive time steps ($s$ is an arbitrary integer). 
Particles in the same platoon are labeled by the same
letter; their phases, not indicated, increase towards the left.
A heavy (red) bar indicates an end-of-platoon.
The particles are in a jammed state (platoons are separated by at most
a single hole).
The particle motion is deterministic except for the exit from site
$L$, which takes place only with probability $\beta$.
Time step $t=s$: Initial configuration. Time step $t=s+1$: The
particle heading platoon $a$ exits and its follower
hops to the exit site.
A new hole, marked by an
up-arrow $\pmb{\uparrow}$, enters the system from the right.
Besides,
platoon $b$ advances by a position exchange with 
the hole marked by a cross $\pmb{\times}$.
Platoon $c$ is blocked.
Time step $t=s+2$: Particle $a$, subject to the random exit rule,
stays on site $L$. Platoons $b$ and $c$ advance.
The two marked holes exchange positions with the platoon to their left.
Time step $t=s+3$: The last particle of platoon $a$ leaves the
system and is replaced with a new hole on site $L$.
Platoon $b$ is blocked but $c$ and $d$ move ahead.
Time step $s=t+4$: Platoons $b$ and $d$ move ahead, but $c$ is blocked.
}
\label{fig_flowL}
\end{figure}

We reason as follows.
A particle on site $L$ has at each time step a
probability $\beta$ to leave the system. The probability $q_r$ that it
exit at the $r$th time step is $q_r=(1-\beta)^{r-1}\beta$, so that for 
leaving the system it needs an average number of time steps equal to
$\sum_{r=1}^\infty r q_r= \beta^{-1}$.
In the jammed state the particle configuration 
consists of platoons. An example is shown in figure \ref{fig_flowL},
where the last particle of each platoon is indicated by a heavy (red)
vertical bar to its left.

When the particle on site $L$ exits, the whole platoon that it was
heading advances one step, so that site $L$ is occupied by the next 
particle in the exiting platoon.
When the last particle of a platoon has left the system, site $L$ is
unoccupied and remains so during one time step, after which it is
occupied by the first particle of the next platoon.
Therefore, the average number of time steps needed for an $n$-particle platoon 
to exit, counted
from the moment that its first particle arrives on site $L$
until the moment that the next platoon's first particle arrives there,
is equal to $n\beta^{-1}+1$.
If $\nu$ is the average platoon length, then 
$\nu\beta^{-1}+1$ is the average time 
needed for $\nu$ particles to exit the system. 
At the right hand end of the lattice we therefore have, 
{\it under jammed flow exit conditions,}

\beq
\Jout = \frac{\nu\beta}{\nu+\beta}\,, \qquad  
\rhoout = \frac{\nu}{\nu+\beta}\,.
\label{xJout}
\eeq 
Before combining our analyses of the entrance and the exit points
we complete this section by finding an expression for the mean platoon
length $\nu$, appearing in (\ref{xJout}).
It should be noticed that $\nu$ is completely determined by the
injection procedure. Actually we shall see that it can be expressed
as a function of $\alpha$.


\subsection{Expression for $\nu(\alpha)$}
\label{sect_nualpha}

The average platoon length $\nu$
depends on the injection procedure defined in
section \ref{sect_enteringrule},
and in particular on the way the phases $\tau_i$
are assigned. Let $\tau_0$ be some arbitrary initial phase.
The phases $\tau_i$ are then related to the
time intervals $T_j$ by
\bea
\tau_i &=& \left[ \tau_0 + \sum_{j=1}^i \left(T_j+1\right) \right]
\,\mbox{mod}\,1\,, \qquad i=1,2,\ldots,
\nonumber\\[2mm]
\tau_i &=& \left[ \tau_0 + \sum_{j=1}^i \hT_j \right]\,
\mbox{mod}\,1
\label{tauhT}
\eea
where we set
\beq
\hT_j \equiv T_j\,\mbox{mod}\,1.
\eeq
The probability distribution $\hP$ for the $\hT_j$
may be calculated as
\bea
\hP(\hT) &=& \int_0^{\infty}\!\dd T\, P(T)\,
             \delta(\hT-T\,{\rm mod}\,1) \nonumber\\[2mm]
&=& \sum_{m=-\infty}^{\infty} \int_0^{\infty}\!\dd T\,
a\ee^{-aT}\,\delta( \hT - T +m ) \nonumber\\[2mm]
&=& \sum_{m=-\infty}^{\infty} a\ee^{-a(\hT+m)}\,
\theta(\hT+m), 
\label{xhatP0}
\eea
in which $\theta(x)$ is the Heaviside step function, here defined by
$\theta(x)=0$ for $x<0$ and $\theta(x)=1$ for $x\geq 0$.
Substituting for $\theta(x)$ its definition in (\ref{xhatP0})
restricts the sum, which can then be calculated as follows
\bea
\hP(\hT) &=& \sum_{m=0}^{\infty} a\ee^{-a(\hT+m)} \nonumber\\[2mm]
&=& \frac{a}{1-{\rm e}^{-a}}\,\ee^{-a\hT}, \qquad 0\leq\hT<1.
\label{xhatP}
\eea
A useful intermediate step is to rewrite (\ref{tauhT}) as
\beq
\tau_i = \htau_i \,\mbox{mod}\,1
\label{taumod}
\eeq
with
\beq
\htau_i = \tau_0 + \sum_{j=1}^i \hT_j.
\eeq
The $\htau_i$ are a sequence of random points on the positive
$\htau$ axis, not confined to [0,1), and there is
at least one point in every interval
between two integers $k$ and $k+1$.
If $\htau_i$ and $\htau_{i+1}$ are in the same interval
$[k,k+1)$, then the corresponding $\tau_i$ and $\tau_{i+1}$ verify
$\tau_i < \tau_{i+1}$ after the modulo in (\ref{taumod}).
If $\htau_i$ lies in $[k,k+1)$ and $\htau_{i+1}$ in $[k+1,k+2)$,
then recalling that $\hT_{i+1} < 1$, we have $\tau_i >\tau_{i+1}$.
Thus each set of $\htau_i$ values within an interval $[k,k+1)$
will give one platoon in the jammed phase.
The average platoon length $\nu$ is therefore equal to the average
number of $\htau_i$ values 
in a unit interval on the $\htau$ axis,
which in turn is the inverse of the average length
of the modulo $1$ time intervals $\hT_j$.
From (\ref{xhatP}) we now obtain easily
\bea
\frac{1}{\nu} &=& \int_0^1\!\dd\hT\,\hT\,\hP(\hT) \nonumber\\[2mm]
&=& 1 + \frac{1}{a} - \frac{1}{1-\ee^{-a}}\,,
\label{xnua}
\eea
which is the desired expression for $\nu$ in terms of $a$,
or equivalently, $\alpha$.

Upon combining (\ref{xnua}) with (\ref{xJout}) we find that the
exiting current $\Jout$ may be expressed as
\bea
\Jout(\alpha,\beta) &=& {\beta}\left[ 1+\frac{\beta}{\nu} \right]^{-1} 
\nonumber\\[2mm]
&=& \frac{\beta}{1+\beta\left(\frac{1+a}{a}\right)
-\left(\frac{\beta}{\alpha}\right)} \nonumber\\[2mm]
&=& \frac{1}{\frac{1+a}{a}+\frac{1}{\beta}-\frac{1}{\alpha}}\,.
\label{xJout2}
\eea
We remark that rewritings like the one above are carried out
most efficiently by working with
$a$ and $\alpha$ as though they were unrelated variables.

This completes the necessary preliminaries.
Everything is in place now for us to determine the phase
diagram of the system subject to the boundary conditions $(\alpha,\beta)$.


\section{Phase diagram}
\label{sect_phase}

In the previous sections, we have identified and characterized
two possible stationary states, free flow and jammed.
It is clear that the free flow state cannot be sustained if the
outflow predicted by (\ref{xJout2})
is smaller than the inflow given by (\ref{xJin}); in that case the
jamming will propagate across the system from exit to entrance.
Inversely, it is also clear that the jammed flow state requires the outflow,
given by (\ref{xJout2}), to be smaller than the inflow 
(\ref{xJin}). Hence in the stationary state viewed as a function of 
$\alpha$ and $\beta$ we expect the occurrence of a phase transition 
on the line
\beq
\Jin(\alpha) = \Jout(\alpha,\beta), 
\label{Jinequality}
\eeq
where we have indicated explicitly the dependence
of the currents on the two model parameters $\alpha$ and $\beta$. 
It is interesting to notice that (\ref{xJout2})
may be rewritten as
\beq
\frac{1}{\Jout} = \frac{1}{\Jin} + \frac{1}{\beta}
- \frac{1}{\alpha}
\label{free_jam}
\eeq
Then one immediately finds the critical line
$\Jin=\Jout$ in 
the $\alpha\beta$ plane is
\beq
\alpha=\beta.
\label{critline}
\eeq
This simple result would have been rather unsurprising
in the case of particle/hole symmetry.
However, in the present case the particle/hole
symmetry is broken by the update scheme, and
we have not found a simpler derivation of (\ref{critline}) than
the one given above.

As a consequence of (\ref{critline}) we have for the current
$J(\alpha,\beta)$ carried by the system in its stationary state
\beq
J(\alpha,\beta) = \left\{
\begin{array}{ll}
\Jin(\alpha),        &\alpha\leq\beta,\\[2mm]
\Jout(\alpha,\beta), &\alpha\geq\beta,
\end{array}
\right.
\eeq
and
\beq
\rho(\alpha,\beta) = \left\{
\begin{array}{ll}
\rhoin(\alpha),        & \alpha<\beta,\\[2mm]
\rhoout(\alpha,\beta), & \alpha>\beta,
\end{array}
\right.
\label{xJ}
\eeq
with $\Jin$ and $\rhoin$ given by (\ref{xJin}) and $\Jout$ and
$\rhoout=\Jout/\beta$ by (\ref{xJout2}).
Across the critical line the current is continuous 
and the particle density jumps by a factor $\beta$,
\beq
\Jinc=\Joutc, \qquad \rhoinc=\beta\rhooutc,
\label{eqJcrhoc}
\eeq
where the subscript `c' indicates evaluation on the critical line.
This jump is a true discontinuity only in the
limit $L\to\infty$. Finite size rounding will be described in the
next section.

The closing remarks of this section 
concern the similarities and differences that appear here 
between the present frozen shuffle update scheme 
and other schemes applied to the TASEP.
First, we note that the critical line (\ref{critline}) 
obtained here for the TASEP
is identical to the one found with other update
schemes, whether it be random sequential update \cite{schutz_d93,derrida93c},
parallel update \cite{degier_n99}, or ordered sequential
update \cite{rajewski98}.
Secondly, however, our expressions for $\Jin$ and $\Jout$ are 
specific to the update procedure adopted here.
In particular, it is remarkable that in the present case
the jammed phase current depends not only
on $\beta$ but also, through the
mean platoon length $\nu(\alpha)$, on the entrance probability $\alpha$.
Indeed, the jammed phase dynamics is completely determined
by the platoon structure, which in turn results from
the injection procedure - whatever this procedure may be.

Thirdly, the frozen shuffle update of this paper, unlike the other
updates, does not lead to a `maximal current' phase:
here the current is always equal to either $\Jin$ or $\Jout$,
that is, always fixed by the boundary conditions.
This feature, confirmed by simulations, 
is due to the deterministic character of the TASEP considered here:
when a particle has the possibility to hop forward,
it does so with probability $1$.


\section{Domain wall picture}
\label{sect_dw}


\subsection{Microscopic domain wall position}
\label{sect_microscopic}

If $\beta < 1$, a waiting
line may be formed at the exit whose rear end is nothing but
a domain wall separating a spatial region with free flow
from one with jammed flow.
For $\alpha<\beta$ the system is in the free flow phase;
the incursions of the jammed region
into the system, starting from the exit point,
will not exceed some finite localization length
and be short-lived. Hence the domain wall is localized near the
exit point.
For $\alpha>\beta$ the system is in the jammed phase;
the waiting line invades
the whole system
and the domain wall is localized near the point of entry.
For $\alpha\approx\beta$ the domain wall position is subject to 
large fluctuations that we will study below; for $\alpha=\beta$
it is completely delocalized.

The frozen shuffle update of the TASEP offers the advantage,
compared to previously studied update schemes,
that the wall position may be defined
very accurately, namely on a microscopic scale%
\footnote{A similar microscopic
definition could be given for a parallel
update with deterministic dynamics.}.
Since in the free flow state no particle is blocked at any time,
it is natural to say that the domain wall position is determined by
the leftmost particle ever to have been blocked during an attempted move.
By convention we will take for the domain wall position $k_{\rm w}$
the half-integer coordinate of
the link on the left of this leftmost blocked particle.
When the leftmost blocked particle
moves again, it carries the wall along;
when another particle positioned
to its left is prevented to move, 
the wall position is transferred to the link on the left of that particle.
All particles to the left of the wall position are in the free flow
state and all those to its right are in the jammed state,
in the sense of the definitions of section \ref{sect_stablestates}.
There is no need to develop tracking strategies
as was necessary \cite{derrida_l_s97},
for example, for the TASEP
with random sequential update.

\begin{figure}
\begin{center}
\scalebox{.45}
{\includegraphics{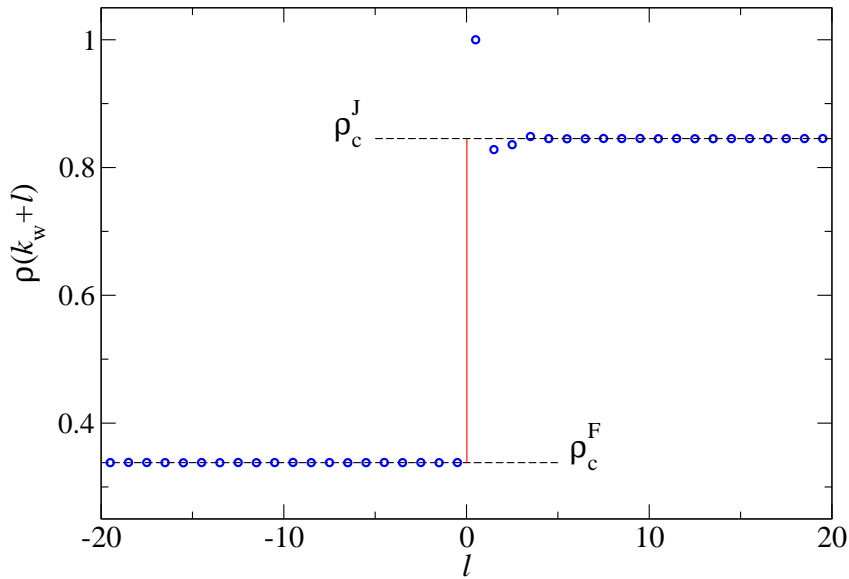}}
\end{center}
\caption{\small 
Density profile $\rho(k_{\rm w}+\ell)$ 
in the coordinate system attached to the domain wall; 
$\ell$ is the distance to the fluctuating wall position $k_{\rm w}$.
The data were taken 
on the critical line with $\alpha=\beta=0.4$.
The profile is very close to a step
function that jumps from $\rhoinc$ to $\rhooutc$ at the vertical (red)
line. Only in the jammed state does there appear some structure
on the scale of a few lattice distances.
}
\label{fig_profile_wall}
\end{figure}

Let $\rho(k)$ be the site dependent particle density.
Figure \ref{fig_profile_wall} shows our simulation results for the
profile in the reference system of the wall,
within an environment of 20 lattice distances around $k_{\rm w}$.
Data were taken only when the wall was more than 20 lattice units
away from both ends.
By construction $\rho(k_{\rm w}+\tfrac{1}{2})=1$.
Apart from this exceptional site the density profile is
almost a step function, separating flat density
profiles with densities $\rhoin$ and $\rhoout$, as shown by the
vertical (red) line in
figure \ref{fig_profile_wall}. Only on the high density side of the
wall does there appear some structure, restricted to a
microscopic distance of the order of a platoon length
(equal to $\nu(0.4)=2.18$ in the case of figure \ref{fig_profile_wall}).

\subsection{Domain wall theory}
\label{sect_domainwall}

\begin{figure}
\begin{center}
\scalebox{.45}
{\includegraphics{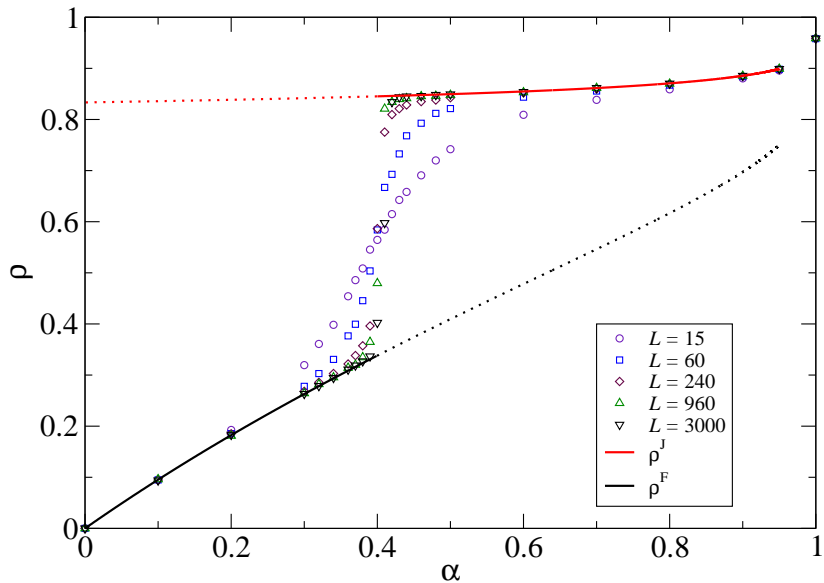}}
\end{center}
\caption{\small Space averaged
particle density $\rho$ versus entrance probability $\alpha$
for the open-ended chain with exit probability $\beta=0.4$.
The simulation data are for a sequence of increasing system sizes. 
They converge towards
the lower (black) and upper (red) solid curves, 
which are the theoretical densities
$\rhoin(\alpha)$ and $\rhoout(\alpha,\beta)$ 
of equations (\ref{xparamrhoin}) and (\ref{xparamrhoout}). 
The dotted curves are their metastable extensions. 
At the critical value
$\alpha=\alpha_{\rm c}=0.4$ the density
jumps from its free flow value $\rhoinc$ to its jammed flow value $\rhooutc$.
The simulation data for finite system sizes $L$, however,
show strong finite size rounding.} 
\label{fig_rhoalpha}
\end{figure}

The strong fluctuations of the wall position near criticality
give rise to large finite size effects visible on the physical observables.
One of these is the spatially averaged density, still denoted $\rho$,
in a finite system. Simulation data for this quantity are shown in
figure \ref{fig_rhoalpha} and are seen to converge only slowly to the
discontinuous theoretical curve.
Below, a description of the system dynamics in terms of the wall motion
will allow us to account for such finite size effects.

Let $Q_k(t)$ be the probability for the wall to be located 
at $k_{\rm w}=k+\tfrac{1}{2}$, {\it i.e.} on link $(k,k+1)$, where
we may take $k=0,1,\ldots,L$.
The ``links'' $(0,1)$ and $(L,L+1)$ are virtual; $Q_0(t)$ is the
probability that the leftmost blocked particle occupies position $k=1$
and $Q_L(t)$ is the probability that none of the particles present in
the system at time $t$ has ever been blocked.   
We will follow here the mesoscopic 
approach to domain wall motion
that was earlier applied with success
to the TASEP with other update schemes \cite{kolomeisky98,pigorsch_s00}
and that has been able to predict
stationary and nonstationary
\cite{dudzinski_s00,nagy_a_s02,santen_a02}
features of those TASEP versions. 
According to this approach we hypothesize that 

(i) the probability $Q_k(t)$ obeys the master equation 
for an asymmetric simple random walk,
\beq
\frac{\dd}{\dd t}Q_k(t) = D_+\left[ Q_{k-1}(t)-Q_k(t) \right]
                        + D_-\left[ Q_{k+1}(t)-Q_k(t) \right],
\label{meq}
\eeq
for $k=0,1,2,\ldots,L$
with the reflecting boundary conditions
$D_+Q_{-1}=D_-Q_0$ and $D_-Q_{L+1}=D_+Q_L$; and

(ii) the coefficients $D_\pm$ in this equation are related to
the currents and the densities in the two phases by
\beq
D_+=\frac{\Jout}{\rhoout-\rhoin}\,, \qquad 
D_-=\frac{\Jin} {\rhoout-\rhoin}\,.
\label{xDpm}
\eeq
For the present case of frozen shuffle update
the validity of equations (\ref{meq}) and (\ref{xDpm}) 
has not been demonstrated on the basis of first principles%
\footnote{That is, starting
from the master equation that defines the time evolution of the particle
configurations.}.
We must therefore consider that they provide
an approximate description.


\subsection{Continuous description}
\label{sect_continuous}

Equation (\ref{meq}) is easily studied on a discrete lattice. However,
when the wall position fluctuates on
the scale of many lattice distances (which it will do
close to the
critical line $\alpha=\beta$), we may replace the index $k$ 
by a continuous variable $0 \leq k \leq L$.  
We will write $Q(k,t)$ instead of $Q_k(t)$
to indicate that we have performed this operation.
Equation (\ref{meq}) then becomes the Fokker-Planck equation
\beq
\frac{\partial Q(k,t)}{\partial t} = 
- \delta\frac{\partial Q}{\partial k}
+ D\frac{\partial^2 Q}{\partial k^2}\,,
\label{FP}
\eeq
in which
\beq
\delta=D_+-D_-\,,    \qquad D=\tfrac{1}{2}(D_++D_-)
\label{dDdelta}
\eeq 
appear as the drift velocity and the diffusion constant, respectively, 
of the domain wall. The boundary conditions become 
$\partial Q(k,t)/\partial k = Q(k,t)/\xi$ for $k=0,L$.
Subject to these, and if we set $\xi=D/\delta$,
the stationary solution of (\ref{FP}) is
\beq
Q_{\rm st}(k)=\frac{\ee^{k/\xi}}{\xi\left( \ee^{L/\xi}-1\right)}\,,
\qquad 0\leq k \leq L.
\label{xQst}
\eeq
For $\xi$ positive (negative) the wall is
localized near $k=L$ (near $k=0$), as anticipated; 
moreover, the domain wall theory gives for the localization length
$|\xi|$ the explicit expression
\beq
\xi=\frac{\Jjam+\Jfree}{2(\Jjam-\Jfree)}
= \frac{2-\Jfree\left(\frac{1}{\alpha}-\frac{1}{\beta}\right)}{2\Jfree
\left(\frac{1}{\alpha}-\frac{1}{\beta}\right)}
\label{xxi}
\eeq
where the second expression was derived with the aid of (\ref{free_jam}).

Let $\rho_L(k)$ denote the particle density at $k$ averaged over
the domain wall fluctuations; we will refer to this function as the
`density profile'.
The domain wall theory gives for the density
profile the following analytic expression,
\beq
\rho_L(k) = \rhoout\int_0^k\!\dd k'\,Q_{\rm st}(k') 
        + \rhoin \int_k^L\!\dd k'\,Q_{\rm st}(k').
\label{defrhok}
\eeq
Working this out with the aid of (\ref{xQst}) we find
\beq
\rho_L(k) = \rhoout + \left(\rhoin-\rhoout\right)
\frac{1-\ee^{-(L-k)/\xi}}{1-\ee^{-L/\xi}}
\label{defrhok2}
\eeq
which is valid for $L\gg 1$ and $k$ of order $L$ (as we used (\ref{xQst}))
and represents an exponential profile
connecting $\rho(0)=\rhoin$ at $k=0$ to $\rho(L)=\rhoout$ at $k=L$.
We recall that
\bea
\rhoin &=& \frac{a}{1+a}\,, \label{xparamrhoin}\\[2mm]
\rhoout &=& \frac{1}{\beta\left( \frac{1}{\rhoin}+\frac{1}{\beta}
-\frac{1}{\alpha} \right)}\, \label{xparamrhoout}
\eea
Equation (\ref{xxi}) shows that
if we vary $\alpha$ (hence $a$) at fixed $\beta$, the localization
length $|\xi|$ diverges at the critical value $\alpha=\alphac=\beta$.
Exactly on the critical line $\alpha=\beta$,
the wall position distribution (\ref{xQst}) becomes flat,
{\it i.e.,}  $Q_{\rm st}(k)=1/L$, the densities on each side of the
wall are related by $\rhoout = \rhoin/\alphac$, and concomitantly
the profile (\ref{xrhok}) becomes the straight line
\beq
\rho_L(k) = \frac{\ac}{1+\ac}
\left[ 1+\frac{1-\alphac}{\alphac}\frac{k}{L} \right],
\label{xrhokc}
\eeq
where $1-\ee^{-\ac}\equiv\alphac$.

\begin{figure}
\begin{center}
\scalebox{.45}
{\includegraphics{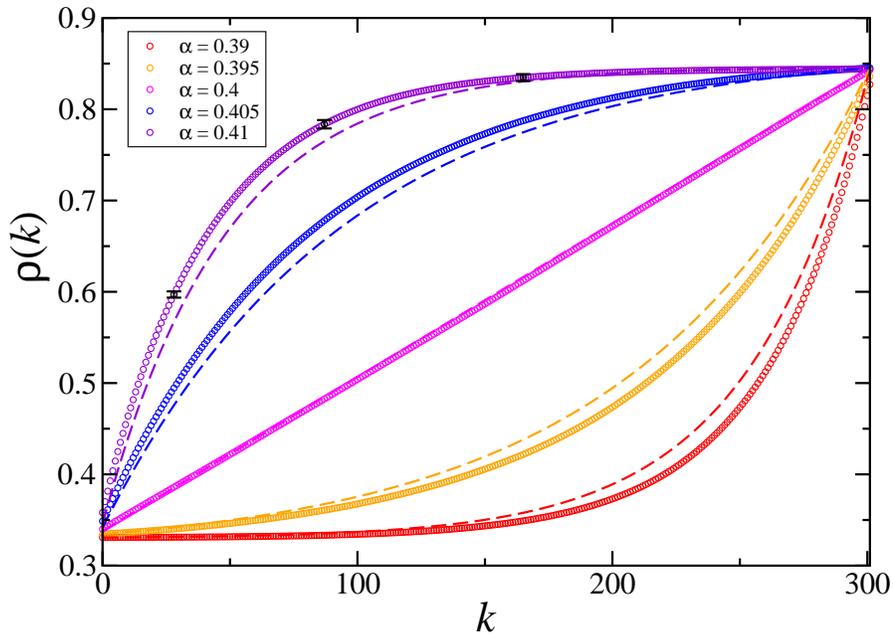}}
\end{center}
\caption{\small 
Spatial density profiles as a function of the lattice site index $k$
in a system of length $L=300$. The
exit probability is $\beta=0.4$ and the entrance probabilities 
$\alpha$ are close to the critical value $\alpha_{\rm c}=0.4$.
The data points join together to produce heavy solid curves.
A few error bars are given for the case
$\alpha = 0.41$. 
For all curves the error bars are of the order of the symbol sizes.
Each simulation had a duration of $8\times 10^7$ time steps.
Dashed curves: the theoretical predictions 
according to equations (\ref{xxi}), (\ref{defrhok2}),
(\ref{xparamrhoin}), and (\ref{xparamrhoout}).
}
\label{fig_profile}
\end{figure}

In figure \ref{fig_profile} we show,
for a system of length $L=302$, for $\beta$ fixed
and for various values of $\alpha$ around $\alphac=\beta$,
the theoretical density profiles $\rho_L(k)$ based on equations
(\ref{xxi}), (\ref{defrhok2}), (\ref{xparamrhoin}), and (\ref{xparamrhoout}).
Also shown are the simulation data for $\rho_L(k)$.
The error bars are of the order of the symbol sizes.
For $\alpha=\alphac$ the simulation results
are indistinguishable from the theoretical straight line (\ref{xrhokc}).
However, for $\alpha\neq\alphac$ there
is a small but definite discrepancy between the theory and
the simulation data. 
We attribute this discrepancy to the approximate nature,
mentioned above, of the master equation (\ref{meq})
on which the theory is based. Finding a theoretical description more
accurate than (\ref{meq}) is at this moment an open challenge.

\subsection{Critical region}
\label{sect_criticalregion}

Near criticality we may expand (\ref{xxi}) in powers of
$\Delta\alpha=\alpha-\alphac$ and obtain to leading order
\beq
\xi \simeq - \frac{\alphac^2}{\Jin \Delta\alpha} = 
- \frac{1+\ac}{\ac}\,\frac{\alphac^2}{\Delta\alpha}\,.
\label{xparamcxi}
\eeq
To lowest order in $\Delta\alpha$ we have
$\rhoout \simeq \rhoin / \alphac$ and thus the
density profile (\ref{defrhok2}) becomes 
\bea
\rho_L(k) & \simeq & \frac{\rhoin}{\alphac} \left[1+(\alphac-1)
\frac{1-\ee^{-(L-k)/\xi}}{1-\ee^{-L/\xi}}\right] 
\nonumber \\[2mm]
& = & \frac{\ac}{(1+\ac)(1-\ee^{-\ac})}\left[1-\ee^{-\ac}
\frac{1-\ee^{-(L-k)/\xi}}{1-\ee^{-L/\xi}}\right]. 
\label{xrhok}
\eea


\subsection{Scaling limit}
\label{sect_scalinglimit}

\begin{figure}
\begin{center}
\scalebox{.45}
{\includegraphics{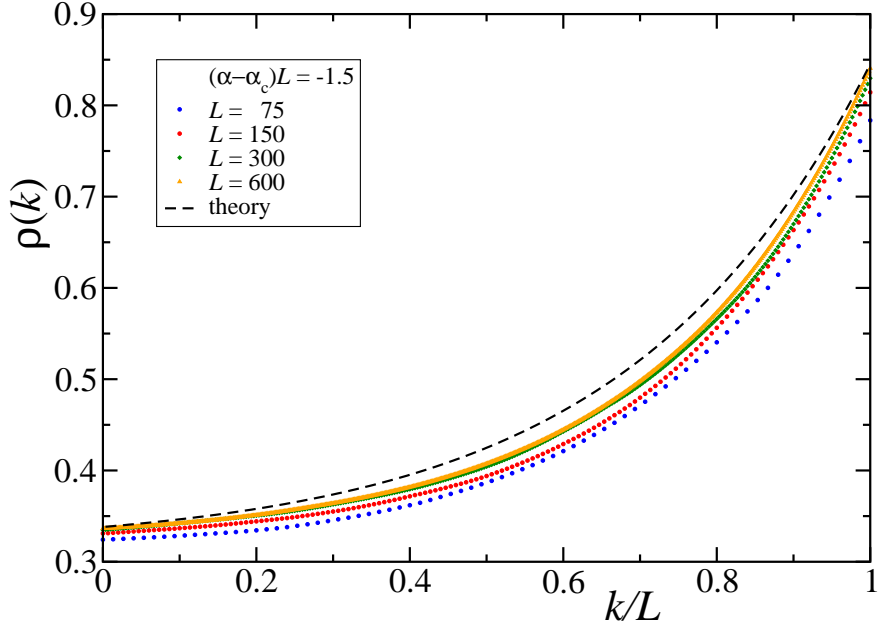}}
\end{center}
\caption{\small 
Spatial density profiles as a function of the scaled lattice site
index $k/L$, obtained from Monte Carlo simulations
for various system sizes $L$ and various $\alpha$ values,
with $(\alpha-\alpha_{\rm c})L=-1.5$ fixed.
In the scaling limit the simulation data converge
to a profile that is seen to depend only on $k/L$, 
as predicted by the domain wall theory.
Dashed curve: the domain wall theoretical prediction for the
density profile in the scaling limit.
The exit probability $\beta=0.4$ is fixed.
}
\label{fig_profilescaling}
\end{figure}

We shall now consider (\ref{xrhok}) in the scaling limit
$\Delta\alpha\to 0$ and $L\to\infty$ with 
$\Delta\alpha L$ fixed.
In this limit the density profile $\rho_L(k)$ becomes a function
of the single ratio $k/L$, as one may see by combining 
(\ref{xrhok}) and (\ref{xparamcxi}),
and noticing that the variable $z$ defined by
\beq
z = 
\lim_{\stackrel{\alpha\to\alphac}{L\to\infty}}\,\, 
\frac{L}{\xi}
  = - \frac{a_c}{1+a_c}\frac{\Delta\alpha\,L}{\alphac^2}
\label{scal_rel}
\eeq
has a fixed value. In this limit the profile (\ref{xrhok}) 
may be simplified to
\beq
\rho_L(k) = \frac{\ac}{1+\ac}
\left[ 1+\frac{1-\alphac}{\alphac}\frac{\ee^{-z(1-k/L)}-\ee^{-z}}{1-\ee^{-z}} 
\right]
\label{xrhokscaling}
\eeq
which depends only on the two scaling variables $z$ and $k/L$. This scaling 
behavior of the profile is confirmed by the simulation data
shown in figure \ref{fig_profilescaling}: for increasing $L$ the
profiles converge to a limit curve. The limiting curve of the
simulation data again
exhibits a slight deviation from the
theoretical curve (\ref{xrhokscaling}).

\begin{figure}
\begin{center}
\scalebox{.45}
{\includegraphics{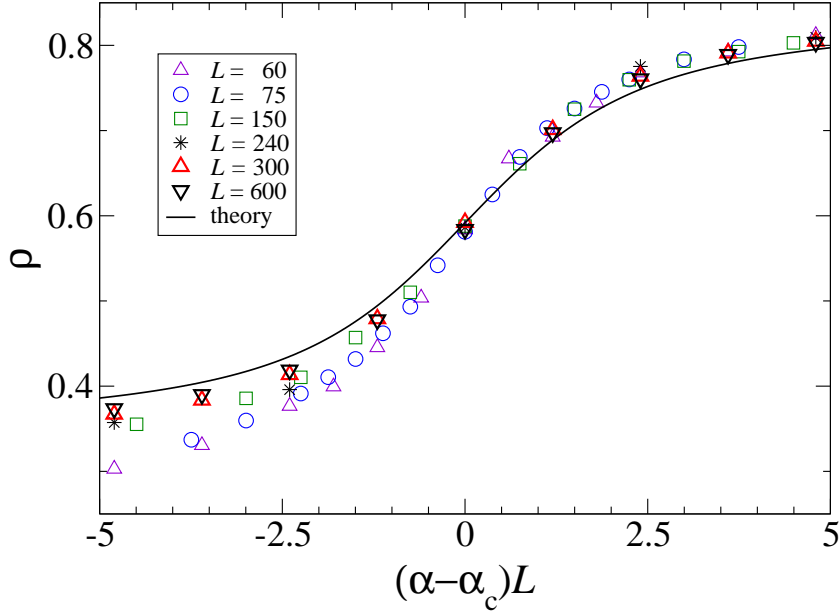}}
\end{center}
\caption{\small 
Space averaged particle density for various system sizes $L$,
as a function of the scaling variable $(\alpha-\alphac)L$.
Solid curve: the theoretical scaling function according to equations 
(\ref{scalingobc2}), (\ref{xPsiz2}), and (\ref{scal_rel}),
which describes the finite size rounding
of the transition seen in 
figure \protect{\ref{fig_rhoalpha}}.
The Monte Carlo data are for $\beta=0.4$, hence $\alpha_{\rm c}=0.4$.
For increasing $L$ they converge to a scaling function whose
deviation from the theoretical prediction is very small
but nevertheless statistically significant. 
}
\label{fig_scalingobc}
\end{figure}

We return now to figure \ref{fig_rhoalpha}
and explain, again in the same scaling limit, the finite size effects
observed there.
In order to obtain the space averaged density we have to determine the
mean value of
the profile (\ref{xrhokscaling}) over $0\leq k/L \leq 1$.
The result is that we obtain for the $L$ and $\Delta\alpha$ 
dependent space averaged
particle density $\rho$ the scaling expression
\beq
\rho = \rhoinc - (\rhooutc-\rhoinc)\Psi(z)
\label{scalingobc2}
\eeq
with $-\infty < z < \infty$ 
and the scaling function $\Psi$ given by%
\footnote{The similarity of this expression to (\ref{xnua}) is
 probably a coincidence.}
\beq
\Psi(z) = 1 + \frac{1}{z} - \frac{1}{1-\ee^{-z}}\,.
\label{xPsiz2}
\eeq
Monte Carlo simulations were carried out on systems of different sizes
$L$. We have seen in figure \ref{fig_profile}
that exactly at the critical point $\alpha=\alphac$
the simulated profile is
linear for all system sizes; as a consequence, in figure 
the space averaged density coincides for $\Delta\alpha L=0$ 
within error bars with the theoretical value.
Within distance $L^{-1}$ from criticality the transient time
necessary to reach stationarity
grows with $L$ and we had to make sure that the data 
represent true stationary state values.
The simulated system sizes $L$ in figure \ref{fig_scalingobc} 
are therefore smaller than the largest values shown in figure
\ref{fig_rhoalpha}.
The Monte Carlo data in figure \ref{fig_scalingobc} confirm
the validity of the scaling as a function of the product variable
$(\alpha-\alpha_{\rm c})L$.

As in figure \ref{fig_profilescaling},
the Monte Carlo curves
in figure \ref{fig_scalingobc}
converge, when $L$ increases,
to a limit curve.
Again, there is a slight but significant
difference between the theoretical limit function 
and the simulation data, a consequence
pointed out above of the fact
that the domain wall theory is only approximate here.

\section{Conclusion}
\label{sect_conclusion}

Motivated by the
modeling of pedestrian motion,
we have in this paper pursued our study 
of a new `frozen shuffle' update scheme for the 
totally asymmetric exclusion process (TASEP) on a one-dimensional lattice.  
An advantage of this scheme is that it fixes
deterministic rules of priority when two hard-core particles
attempt to hop to the same target site,
a feature that allows for more efficient Monte-Carlo simulations.
This is in contrast with the widely used parallel update
which, when two pedestrians want to hop
to the same site, generates conflicts
which have to be solved by extra ad-hoc rules.
Actually these algorithmic conflicts
may have a physical counterpart:
In dense crowds conflicts 
between real interacting pedestrians
are thought to be one of the sources for the
clogging effect observed near exits in
emergency evacuation \cite{kirchner_n_s03}.
However, it is still an open question to determine
whether the effect of the conflicts due to parallel update
is overestimated or not, compared to the effect of
real-life conflicts between pedestrians.
Shuffle update yields an alternative that 
solves conflicts in a smoother way and may be more
appropriate for pedestrian modelling
in particular when the density is not too high.
Besides, this scheme is easily
accessible to analytical study.

Compared to other update schemes,
the frozen shuffle update modifies the properties of the stationary
state in the closed system studied
in reference \cite{appert-rolland_c_h11a}
as well as in the open system studied here.
The open system, just like the closed one, appears capable
of sustaining `free flow' and `jammed' states.
More specifically, the jammed state's behavior is
fully driven by the platoon structure, where platoons
are ensembles of neighbor particles that move as a whole at each
time step.
As the platoon structure depends on the injection procedure,
the flux of the jammed phase for the frozen shuffle update
(in contrast to other update schemes) depends not only
on the exit probability $\beta$ but also on the entrance
probability $\alpha$.

In spite of this specificity, the transition line
between the free flow and the
jammed phase is found to be the usual $\alpha=\beta$ line.
As we considered only the deterministic version of the
model (particles move forward with probability $1$
when the target site is empty), there is no maximal
current phase.
It could be of interest to extend the study to the
case of particles hopping with probability $p<1$.

We have accompanied our theoretical results by Monte Carlo simulations.
and found excellent agreement for the phase diagram.

Finally, in order to explain
the finite system size effects observed in the Monte Carlo
simulations and especially important near the transition, 
we have applied a domain wall approach 
originally developed in references 
\cite{kolomeisky98,pigorsch_s00,dudzinski_s00,nagy_a_s02,santen_a02}.
Frozen shuffle update with deterministic dynamics
leads to the remarkable property
that the position of the domain wall that separates two homogeneous
regions
can be determined
at the microscopic scale of the lattice distance.
The domain wall theory appears to account quite well
for the finite size effects, but we point out that
nevertheless a small but definite discrepancy persists
between this theory and the simulation data.
The discrepancy concerns in particular the wall
localization length, near the border of the system,
for reasons still to be understood.

An update scheme related to ours and preceding it, called
`random shuffle' update, has been proposed \cite{wolki_s_s06,smith_w07a}
for the improvement of pedestrians models.
That scheme has no obvious interpretation
in terms of the interaction between pedestrians and
should rather be viewed as an algorithm devised
just to avoid conflicts.
By contrast, the `frozen shuffle' variant presented
in this paper has a physical interpretation
in terms of phase differences in the stepping cycle,
which makes it relevant for applications to pedestrians.

In forthcoming work \cite{appert-rolland_c_h11c}
we will study the case of two intersecting lanes
of pedestrians (TASEPs) simulated with frozen shuffle update.


\vspace{2cm}

\end{document}